\documentclass[aps,pra,showpacs,twoside,twocolumn,10pt]{revtex4-1}
\usepackage[colorlinks=true, citecolor=red, urlcolor=blue ]{hyperref}
\usepackage{epsfig,newlfont,amssymb,amsfonts,amsmath,bm,subfigure,palatino,mathtools,amsthm,braket,soul,enumitem,color,graphics,graphicx,times,physics,bbold}
\usepackage[normalem]{ulem}
\usepackage{xcolor}
\usepackage{physics}
\usepackage{dsfont}
\usepackage{mathrsfs}
\usepackage{verbatim}
\usepackage{amsthm,amssymb}
\usepackage{amsmath}
\usepackage{float}

\begin{document}
\title{
Advancing Quantum Otto Engine Performance via Additional Magnetic Field and Effective Negative Temperature
}

\author{Arghya Maity, Aditi Sen(De)}
\affiliation{Harish-Chandra Research Institute, A CI of Homi Bhabha National
Institute,  Chhatnag Road, Jhunsi, Allahabad - 211019, India}

\begin{abstract}

We formulate a protocol for a four-stroke quantum Otto engine that is capable of achieving superior performance when operating between two thermal reservoirs: one at a positive spin temperature and the other at an effective negative spin temperature.
We adopt a protocol that encompasses a rotating magnetic field in the \((x, y)\)-plane, as well as an additional magnetic field in the \(z\)-direction that possesses distinct strengths. Consequently, we acquire the capability to manipulate the  strength of the magnetic field autonomously in both directions during dynamics. We report that by precisely adjusting the strength and the direction of the magnetic field in the \(z\)-direction and manipulating other relevant system parameters, we can effectively enhance the transition probability and hence the efficiency of the engine as well which can not be achieved without the additional magnetic field, although the impact is not ubiquitous. Additionally, another important significance of our model is that these engines operate within an extended operational domain, reaching into temperature ranges where 
 the effective negative temperature-based quantum Otto engines operating only on the rotational magnetic field in the \((x,y)\) plane, are unable to function.
Specifically, we identify a threshold value for the magnetic field, dependent on the driving time, at which an improvement in efficiency is observed.
We propose that this advantage may arise from the system exhibiting greater coherence with respect to the driving time, which we evaluate using the $l_1$-norm coherence measure. 
Another noteworthy aspect is that the advantage in efficiency gained from the additional  magnetic field continues to surpass that of the protocol without the field, 
even in the presence of impurities in the magnetic field having  a specific range of disorder strengths. 
 
\end{abstract}

\maketitle

\section{Introduction}

The goal of a thermal machine is to maximize the work which is obtained from the conversion of heat through a reversible process \cite{Francisco_Entropy_2020}. 
In the middle of last century, it was discovered that the efficiencies of  thermal machines can be qualitatively improved if they are built using quantum systems having finite or infinite dimensions \cite{Scovil_PRL_1959, Gemmer_QT_book, Sebastian_book, Sebastian_Review_AVS_2022}. 
By exploiting atomic coherence, a quantum Carnot engine containing a heat bath of three-level atoms was found which can provide higher efficiency than that can be obtained by classical engine \cite{Scully_Science_2003, Scully_Science_2003}.
Heat engines are typically designed to operate between two equilibrium reservoirs although non-equilibrium reservoirs (engineered by varying external parameters) can also drive heat engines. Therefore, nonequilibrium reservoirs are often referred to as engineered reservoirs, especially when they are created or manipulated by varying external parameters to achieve specific non-equilibrium conditions. The presence of resources like quantum coherence  \cite{Scully_Science_2003}, quantum correlations~\cite{Dillenschneider_EPL_2009, Llobet_PRX_2015}, and squeezed thermal reservoirs~\cite{ Scully_PRL_2001, Yanchao_PhysicaA_2020, Rafael_PRB_2022, Abah_PRL_2014, Jan_PRX_2017} in these heat engines is responsible for larger  efficiency  than the Carnot efficiency  although this does not violate the second law of thermodynamics (see also Refs. \cite{Niedenzu_Nature_2018, Dillenschneider_EPL_2009, Huang_PRE_2012, Abah_EPL_2014, Abah_PRL_2014, Hardal2015, Niedenzu_2016, Manzano_PRE_2016, Jan_PRX_2017, Bijay_PRB_2017}). 
Furthermore, heat engines have been studied with different reservoirs including squeezed \cite{Jan_PRX_2017, Assis_JOP_2021, Yanchao_PhysicaA_2020, Rafael_PRB_2022, Hardal_PRE_2017, Scully_PRL_2001, Huang_PRE_2012, Zagoskin_PRB_2012, Abah_PRL_2014, Ferdi_PRE_2014, Rui_PRE_2015, Xiao_PLA_2018, Niedenzu_Nature_2018, Wang_PRE_2019, Assis_PRE_2020}, spin \cite{Jackson_PRA_2018}, superconducting reservoirs \cite{Rafael_PRB_2022, Hardal_PRE_2017}, etc. Importantly, current technological advances promise to build such reservoirs in physical systems like trapped ions, cavity QED, nuclear magnetic resonance  etc \cite{Murch_Nature_2013, Carvalho_PRL_2001, Werlang_PRA_2008, Pielawa_PRL_2007, Werlang1_PRA_2008, Prado_PRL_2009, Verstraete_Nature_2009, Hama_PRL_2018, Lovric_PRA_2007, Alvarez_PRL_2011}. The main motive behind this quantum reservoir engineering is to find ways to obtain a better overall performance of the engines \cite{Sebastian_Review_AVS_2022, Mendon_PRR_2020}. On the other hand, there are also constant efforts to quantify quantum correlations from the thermodynamic perspective \cite{ollivier2001, horodecki2002, aditi2005, bera2017} and such quantities can also be shown to have relation with the efficiency of quantum Carnot engines \cite{Dillenschneider_EPL_2009, Llobet_PRX_2015}.

In the practical world, the  Otto cycle is used in a spark-ignition engine, which is commonly found in cars and small vehicles \cite{Breeze_book_2019}. 
The Otto cycle was developed for a number of reasons, including great efficiency while emitting little emissions \cite{Heywood_book_2018, Philip_book_1991, Stone_book_1985}, and has applications  ranging from small generators to massive industrial engines.
In a similar spirit, it was realized that the quantum version of the Otto engine (QOE) consists of four strokes -- in two of the strokes, a working medium is connected to cold or hot thermal reservoirs while the working medium either expands or compresses unitarily, thereby producing work in the other two strokes.  
In this work, we will be concentrating on two thermal reservoirs, one at a positive temperature while the other one is prepared at an effective negative temperature \cite{Assis_PRL_2019, Jens_PRX-Quantum_2022} based on the population of energy eigenstates. Recently, it was demonstrated experimentally in a nuclear magnetic resonance (NMR) set-up that if both the reservoirs are at positive temperatures, the work of QOE cannot defeat the corresponding classical  Otto engine (COE) \cite{Peterson_PRL_2019} while a higher efficiency than that of the COE is obtained if one of the reservoirs is at effective negative temperature \cite{Assis_PRL_2019}. 
At this point, one can ask following  questions which naturally arise both from the theoretical and experimental perspectives.   (1) What is  the most general quantum Otto engine  model that can deliver benefits when one of the reservoirs  is engineered with effective negative temperature?
 (2) Secondly, what properties of the state which are generated during the cycles are responsible for providing such advantages?
(3) Thirdly, is the QOE robust against any kinds of impurities which typically arise during implementation? 

In this article, we introduce a design of QOE in which the driving Hamiltonian is considered in its comprehensive form, thereby answering the first question. 
In particular, we deal with a rotating magnetic field within the \((x,y)\)-plane, and an additional magnetic field in the \(z\)-direction where the strengths of the magnetic fields are distinct in two situations.
As a result, we gain the ability to manipulate the strength independently in both the directions, thereby increasing the complexity of the system. 
Due to this,  the calculation of efficiency in this scenario is comparatively challenging than the OQE without the magnetic field in the \(z\)-direction. 
We manage the unitary evolution by expressing it on a rotating basis that allows us to investigate the transition probability as a function of the driving time for varying strengths of the magnetic field in the \(z\)-direction.

We report that the efficiency of QOE increases with the increase in the strength of the magnetic field in the \(z\)-direction. However, such enhancement is not ubiquitous. It depends on the driving time and effective negative temperature. In particular, with the increase in driving time period, the enhancement fades off. Moreover, the enhanced transition probability leading to the improvement in efficiency due to the additional magnetic field becomes apparent only beyond a certain threshold value of the population of the excited states. Moreover, a crucial aspect of our model lies in its ability to function within an expanded operational domain. It extends into temperature ranges where the  effective negative temperature-based quantum Otto engine, which operates solely on rotational magnetic fields in the \((x,y)\) plane, is incapable of operating. Fascinatingly, with the aid of an additional magnetic field in the \(z\)-direction, the protocol demonstrates the capacity to function as an engine across the entirety of the effective negative temperature spectrum.\\
We further assert that the dynamics of coherence during the expansion and compression steps play a significant role in achieving high efficiency in the presence of negative temperature. Specifically, we discover that the states exhibit increased coherence based on the $l_1$-norm measure \cite{Streltsov_RMP_2017} when the strength of the magnetic field in the 
\(z\)-direction is moderate compared to the scenario without the magnetic field. Therefore, observation identifies the resource which is responsible for obtaining the higher efficiency in QOE. In order to visualize the enhancement in the proposed protocol, we select all the parameters involved in this scheme with the recent experiment on QOE \cite{Assis_PRL_2019}. Notice that the QOE setup combined in Ref. \cite{Assis_PRL_2019} is a subclass of the results obtained in this current manuscript.

 Execution of each step perfectly in QOE  is an ideal scenario. It is natural that imperfections enter \cite{Maciej_AIP_2007, Ahufinger_PRA_2005, Tanoy_PRA_2022, Ahana_PRA_2020, Srijon_PRA_2020, Maciej_Nature_2010} in the engine during the preparation of thermal states or during unitary dynamics. Assessing the robustness against disorder becomes a crucial consideration in the design of quantum devices which is one of the aims of this work. We introduce disorder into the ideal engine parameters by randomly selecting them from Gaussian and uniform distributions. Subsequently, we demonstrate the remarkable robustness in the efficiency of the quantum Otto engine based on effective negative temperature against such impurities. The level of robustness strongly relies on the state populations, with higher effective negative temperatures exhibiting increased resilience. This phenomenon holds great promise not only in the field of quantum engines but also in the implementation of effective negative temperature in other quantum devices.
Remarkably, our findings reveal that the advantage in efficiency achieved through an additional magnetic field in the presence of the disorder is higher than that can be obtained without the magnetic field in the \(z\)-direction in the absence of any impurities present in the system.

In Sec. \ref{four-strokes}, we provide the framework of four-stroke quantum Otto engine. In Sec. \ref{transition_prob}, the transition probability is analytically calculated and the nature of transition probability is also shown graphically.   We compute the efficiency and coherence in Sec. \ref{transition_prob} and try to argue that they are interconnected.
Sec. \ref{disorder} considers the effects of disorder on the performance of the engine. A conclusion is presented in Sec. \ref{conc}.

\section{Design of Quantum Otto engine}
\label{four-strokes}

\begin{figure}
\includegraphics[width=\linewidth]{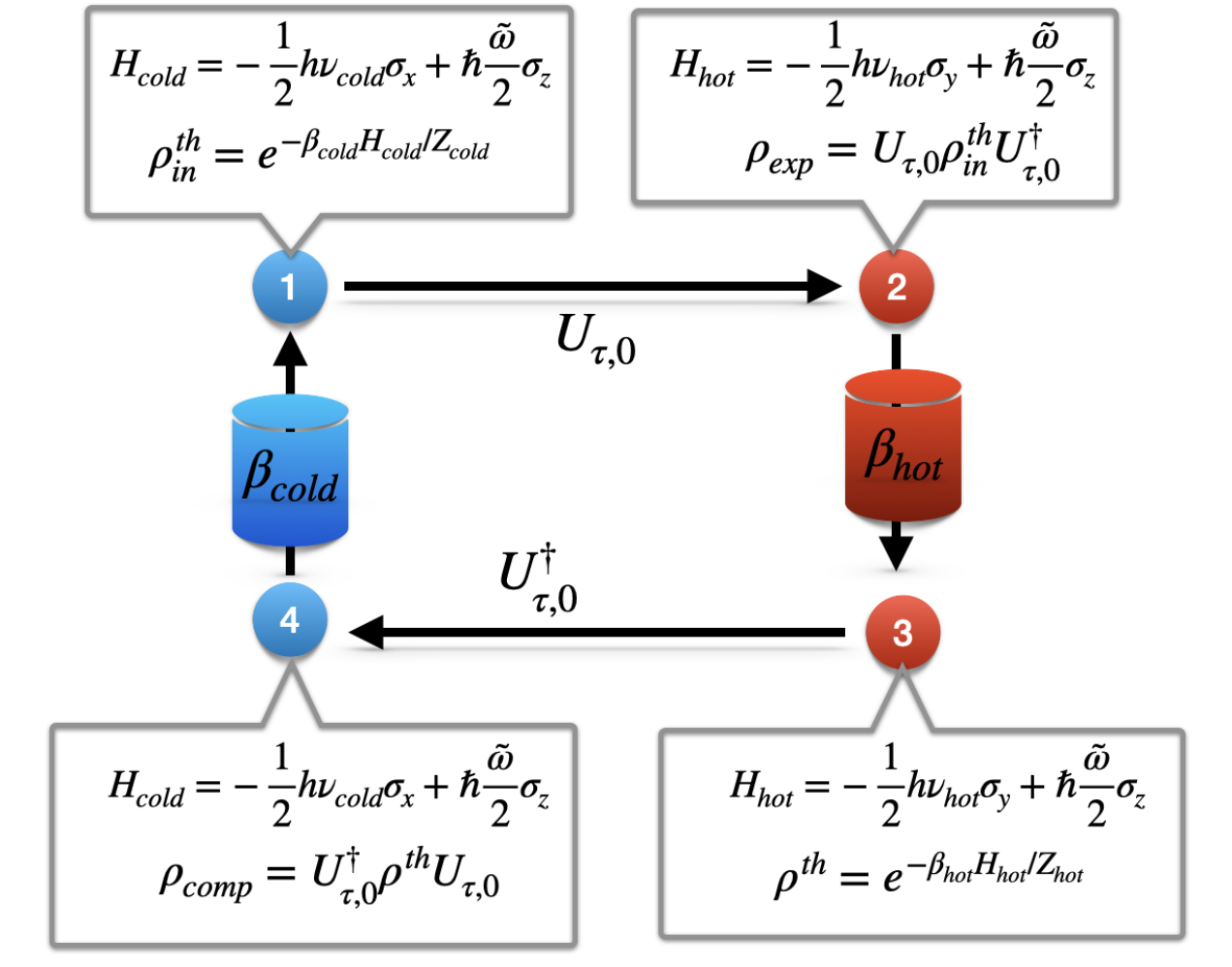}
\caption{Schematic diagram of four-stroke Otto cycle. 
 The path $1\rightarrow 2$ indicates the expansion stroke, $2\rightarrow 3$ heating stroke, while  $3 \rightarrow 4$ represents the compression stroke and $4\rightarrow 1$  is the cooling stroke. States and the corresponding Hamiltonian  in each stroke  are mentioned. Note that $1\rightarrow 2$ and $3 \rightarrow 4$ evolve  unitarily.     }
\label{fig:Otto_cycle_diagram}
\end{figure}

A four-stroke quantum version of a Otto engine operates between two thermal reservoirs -- one of them is at positive spin temperature while the other one is at effective negative spin temperature \cite{Jens_PRX-Quantum_2022, Assis_PRL_2019}. Typically, the four strokes of QOE consists of
(i) cooling, (ii) expansion, (iii) heating, and (iv) compression  (see Fig. \ref{fig:Otto_cycle_diagram} for the schematic diagram). Note that in each stroke, either heat or work is exchanged but never both are converted simultaneously.

\textbf{1. Cooling stroke.} 
For a given Hamiltonian, $H_{cold}$,
the working medium is initially prepared in a thermal state, i.e., $\rho^{th}_{in} = e^{-\beta_{cold}H_{cold}}/Z_{cold}$, where  $\beta_{\text{cold}} =1/{k_B T_{\text{cold}}}$ with $k_B$ being the Boltzmann constant and $T_{\text{cold}}$ being the cold effective spin temperature, and $Z_{\text{cold}}= \Tr[e^{-\beta_{cold}H_{cold}}] $ represents the partition function. 
Unlike the previous works \cite{Assis_PRL_2019}, the Hamiltonian corresponding to the  equilibrium state can be represented as $H_{\text{cold}} = -\frac{1}{2} h \nu_{\text{cold}} \sigma_x + \hbar \frac{\Tilde{\omega}}{2} \sigma_z $ where $h$ is the Planck constant, $\hbar=\frac{h}{2\pi}$ and $\nu_{\text{cold}}$ is the frequency, fixed by the physical system where the experiment can be performed, $\sigma_{i}$ \((i=x,y,z)\) represents the Pauli matrices, and $\Tilde{\omega}$ is the strength of the magnetic field in the z-direction whose details is described below. 
\\

\textbf{2. Expansion stroke.}  
The system evolves from time, $t=0$ to $t=\tau$ according to the Hamiltonian,
\begin{equation}
\label{Equ:driven_Hamiltonian}
    H_{exp}(t) = -\frac{1}{2}h\nu(t)\Big[\sigma_x \cos wt  + \sigma_y \sin wt \Big ] + \hbar \sigma_z \frac{\Tilde{\omega}}{2}, 
\end{equation} 
where $\omega=\frac{\pi}{2\tau}$, thereby ensuring a full rotation of the magnetic field from $x$- to $y$-direction \cite{Peterson_PRL_2019, Denzler_PRR_2020}, $\nu(t) = \nu_{cold}\Big[ 1- \left (\frac{t}{\tau} \right )\Big ] + \nu_{hot} \left (\frac{t}{\tau} \right )$ which reflects how the energy spacing widens from $\nu_{cold}$ at time $t=0$ to $\nu_{hot}$ at $t=\tau$, and additionally, we consider a constant magnetic field with strength $\frac{\Tilde{\omega}}{2}$ and $\Tilde{\omega} = \mathbf{g} \omega$ with $\mathbf{g}$ being a constant. The final evolution time, $\tau$, is fixed by the experiment performed. Moreover, $\tau$ should be chosen to be much shorter than the decoherence time,  so that the evolution can be performed via unitary operator \cite{T.B.Batalhao_PRL_2014, T.B.Batalhao_PRL_2015}. Note that the driving Hamiltonian which has only rotating magnetic field with varying strength $\nu(t)$ in the \((x,y)\)-plane are used to design quantum heat engines \cite{Peterson_PRL_2019, T.B.Batalhao_PRL_2014, T.B.Batalhao_PRL_2015, Hiromichi_Entropy_2023, Denzler_PRR_2020}. We will demonstrate that the constant magnetic field, associated with the rotating field can give rise to a significant difference in the performance of QOE.
\\

The unitary operator responsible for the dynamics, in this case, takes the form as $ U_{\tau,0} = \mathcal{T} e^{-\frac{i}{\hbar} \int_{t=0}^{\tau} H_{exp}(t) dt}$ where $\mathcal{T}$ is the time-ordering parameter. At the end of the expansion stroke, i.e., at time $\tau$, the Hamiltonian reduces to $$
H_{hot}= -\frac{1}{2} h \nu_{hot} \sigma_y + \hbar \frac{\Tilde{\omega}}{2} \sigma_z ,
$$ 
since at \(t=\tau\), \(\nu(t= \tau) = \nu_{hot} \) while the resulting state becomes $\rho_{exp} = U_{\tau,0} \rho^{th}_{in} U^{\dagger}_{\tau,0}$.  
The transition probability ($\xi$) between the eigenstates of the Hamiltonians $H_{cold}$ and $H_{hot}$ is defined as 
\begin{equation}
\label{Equ:transitin_prob}
    \xi = | \bra{\Psi^{hot}_{\pm}}U(t)\ket{\Psi^{cold}_{\mp}} |^{2} =  | \bra{\Psi^{cold}_{\pm}}U^{\dagger}(t)\ket{\Psi^{hot}_{\mp}} |^{2} ,
\end{equation}
where $\ket{\Psi^{hot}_{\pm}}$ and $\ket{\Psi^{cold}_{\mp}}$  are the eigenstates of $H_{hot}$ and $H_{cold}$ respectively.
When the process obeys the adiabatic theorem and there is no transition between the instantaneous eigenstates of the Hamiltonian, $\xi$ vanishes. Due to the transition between the instantaneous eigenstates of the Hamiltonian (i.e., associated with a finite time), the irreversibility  is introduced in the QOE  which leads to the low performance for general quantum engine protocol when both the thermal reservoirs are at a positive spin temperature \cite{Peterson_PRL_2019, Plastina_PRL_2014, Cakmak_EPJD_2017}. Interestingly, when one considers two reservoirs having a positive and an effective negative spin temperatures, the performance of the QOE is enhanced \cite{Assis_PRL_2019}.

\textbf{3. Heating stroke.} 
This is again a thermalization process (see $2 \rightarrow 3$ step in Fig. \ref{fig:Otto_cycle_diagram}) in which the state reaches to the equilibrium state corresponding to the Hamiltonian, \(H_{hot}\), represented as $\rho^{th} =e^{-\beta_{hot}H_{hot}}/ \Tr[e^{-\beta_{hot}H_{hot}}]$. It occurs through the heat exchange between the working medium and the bath.
\\

\textbf{4. Compression stroke.} 
It is the fourth stroke of the Otto cycle and is the reverse process of the expansion in which an energy gap compression is attained \cite{Camati_PRA_2018, Campisi_RMP_2011, Tien_Kieu_PRL_2004}.
Therefore, time-reversed protocol of the expansion stroke occurs such that the Hamiltonian can be written as $H_{comp}(t)=-H_{exp}(\tau - t)$. As in the case of expansion, it can also be assumed to be unitary, denoted as $U^{\dagger}_{\tau,0}$. The compression stroke acts on $\rho^{th}$ and leads to the final state as $\rho_{comp} = U^{\dagger}_{\tau,0}\rho^{th} U_{\tau,0}$ in which the Hamiltonian can be written as 
$$
H_{cold} = -\frac{1}{2} h \nu_{cold} \sigma_x + \hbar \frac{\Tilde{\omega}}{2} \sigma_z
$$
 since at $t=0$, $\nu({t=0})=\nu_{cold}$.

\section{Enhanced Efficiency via constant magnetic field}
\label{transition_prob}

\subsection{Transition Probability}
\begin{figure}[h!]
\includegraphics[width=\linewidth ]{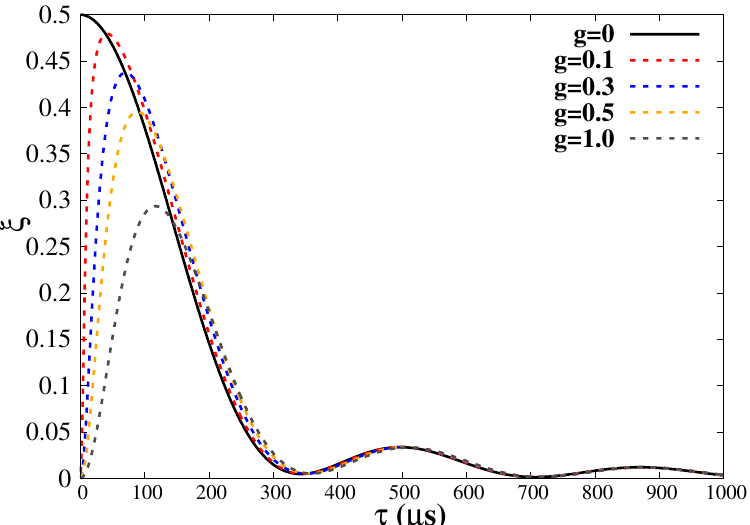}
\caption{Transition probability ($\xi$) (ordiante) with respect to driving time $\tau$ (abscissa) for different  strengths of the magnetic field in the \(z\) direction  ($\mathbf{g}$). The driving time is taken in microseconds and the ordinate is dimensionless. We take $\nu_{cold} = 2.0$ and $\nu_{hot}= 3.6 kHz$. For  $\mathbf{g}=0$, $\xi$ is plotted in solid line and for higher $\mathbf{g}$ values, it is plotted in dashed lines. }
\label{fig:transitin_prob}
\end{figure}

To establish the role of additional magnetic field in the performance of QOE, let us now investigate the effects of $\mathbf{g}$ in \(\Tilde{\omega}\) on the transition probability and the efficiency  of QOE. To obtain the transition probability, $\xi$ in Eq. (\ref{Equ:transitin_prob}), we use the eigenvalue equation of $H_{cold(hot)}$ corresponding to the eigenstate $\ket{\Psi^{cold(hot)}_{\pm}}$ as  
\begin{eqnarray}
\label{Equ:Eigenvalue_H}
     && H_{cold(hot)} \ket{\Psi^{cold(hot)}_{\pm}} \nonumber \\
    &&=\pm \frac{h}{4\pi} \sqrt{4\pi^2 \nu^{2}_{cold(hot)}+ \Tilde{\omega}^{2} } \ket{\Psi^{cold(hot)}_{\pm}} \nonumber \\
    && =\pm E_{cold(hot)} \ket{\Psi^{cold(hot)}_{\pm}}
\end{eqnarray}
where $E_{cold(hot)}= \frac{h}{4\pi} \sqrt{4\pi^2 \nu^{2}_{cold(hot)}+ \Tilde{\omega}^{2} } $.
Let us rewrite the Hamiltonian for the dynamics in step 2, i.e.,
\begin{eqnarray}
\label{Equ:hamiltonian}
H_{exp}(t) = && -\frac{1}{2}h\nu(t)\Big[\sigma_x \cos wt +\sigma_y \sin wt \Big ] + \hbar \sigma_z \frac{\Tilde{\omega}}{2}  \nonumber \\
  = &&  -\frac{1}{2}h\nu(t)\Big[(\sigma_{+}+\sigma_{-}) \cos wt +  (\sigma_{+}-\sigma_{-}) \sin wt   \Big ] \nonumber \\
     && ~ ~ ~ ~ ~ ~ ~ ~ ~ ~ ~ ~ ~ ~ ~ ~ ~ ~ ~ ~ ~ ~ ~ ~ ~ ~ ~ ~ ~ ~ ~ ~ ~ ~ ~ ~ ~ ~ ~ ~ ~ ~ ~ ~ ~ ~ ~ ~ ~ ~ ~ +  \hbar \sigma_z \frac{\Tilde{\omega}}{2}  \nonumber \\
  = && -\frac{1}{2}h \nu(t)\Big[e^{-i\frac{\omega}{2}\sigma_z t} \sigma_x e^{i\frac{\omega}{2}\sigma_z t}  \Big ] + \hbar \sigma_z \frac{\Tilde{\omega}}{2} ,
\end{eqnarray}
where $\sigma_{\pm} = \frac{1}{2}\left(\sigma_{x} \pm i \sigma_{y}  \right )$.
Now from the equation, $i \hbar \frac{dU(t,0)}{dt} = H_{exp}(t)U(t,0)$ for the Hamiltonian of above Eq. (\ref{Equ:hamiltonian}) we can get

\begin{eqnarray}
    i \frac{dU^{\prime}(t,0)}{dt}  =  \left[  - \pi \nu(t) \sigma_x + \left( \frac{\Tilde{\omega}}{2} - \frac{\omega}{2}\right)\sigma_z \right ] U^{\prime}(t,0) ,
\end{eqnarray}
where $ U^{\prime}(t,0) = e^{i\frac{\omega}{2}\sigma_z t} U(t,0) $.
If we consider the same magnetic strength for rotating axis and z-axis, i.e, $\Tilde{\omega}=\omega$, it is very straight forward. In this article, we are curious about the the case when $\Tilde{\omega} \neq \omega$. The principle question that we want to investigate is whether there is any advantage of taking $\Tilde{\omega}$ instead of taking $\omega$.  
\\
We can write the time-evolution operator corresponding to Hamiltonian as \cite{Barnes_PRL_2012, Barnes_PRA_2013}
\begin{equation}
U=
    \begin{pmatrix}
u_{11} & -u^{*}_{21}\\
u_{21} & u^{*}_{11}
\end{pmatrix}, ~ ~ ~ |u_{11}|^{2} + |u_{21}|^{2}=1, 
\end{equation}
and we transform the unitary to a rotating x-basis. We have
\begin{equation}
    D_{\pm} = \frac{1}{\sqrt{2}} e^{\pm i J(t)} \{ u_{11}(t) \pm u_{21}(t)  \} 
\end{equation}
where $J(t) = \int_{0}^{t} - \pi \nu (t^{\prime})dt^{\prime} $.
By solving the Schr$\ddot{o}$dinger equation, we get
\begin{equation}
\label{Eq:1st_order}
    \dot{D}_{\pm}(t) = -i \left( \frac{\Tilde{\omega}}{2} - \frac{\omega}{2}\right ) e^{\pm 2iJ(t) } D_{\mp}(t) .
\end{equation}
For the Hamiltonian where $\Tilde{\omega}=\omega$, and from this first order differential equation, we can construct the unitary which alike with the unitary of Ref. \cite{Denzler_PRR_2020}. For our case, we go to the second order differential equation for $D_{+}$ which is
\begin{equation}
    \ddot{D}_{+}(t) + 2i\pi \nu(t) \dot{D}_{+}(t) + \left( \frac{\Tilde{\omega}}{2} - \frac{\omega}{2}\right )^{2} D_{+}(t)=0.
\end{equation}
By solving the above differential equation, we can obtain the unitary operator,  $U_{\tau,0}$  which leads to the transition probability $\xi$ between the energy eigenstate of $H_{cold}$ and $H_{exp}$ during the compression stroke. The transition probability, $\xi$, with driving time, $\tau$ is shown in Fig. \ref{fig:transitin_prob}.
\\
\\
Unlike the conventional Hamiltonian considered in QOE (without constant magnetic field in the \(z\)-direction), $\xi$  increases with the expansion (compression) time, \(\tau\) and then decreases with $\tau$ as is the case for Hamiltonian with $\Tilde{\omega}=0$. For a certain nonvanishing $\mathbf{g}$ values, we clearly find that there exists a range of driving time, $\tau$, such that 
$\xi(\mathbf{g}>0) > \xi(\mathbf{g}=0)$ (see Fig. \ref{fig:transitin_prob}). It is also noted that such advantage decreases with the increase of $\mathbf{g}$ and high value of $\tau$. Notice that the net work of the cycle, and hence the efficiency of QOE is related to $\xi$. Therefore, we can expect that if we choose $\tau$ close to the case with $\xi$ being maximum or higher than the one obtained with $\mathbf{g}=0$, the efficiency can be enhanced. We will now manifest that this is indeed the case. For demonstration, we choose those set of parameters which are used in NMR experiment \cite{Assis_PRL_2019, Peterson_PRL_2019} so that the benefit of additional magnetic  field can be illustrated. 

 \subsection{Calculate the Efficiency}
 \label{efficiency}
Before computing the net work and efficiency, let us define the local temperature in terms of population as
\begin{equation}
\label{Equ:relation_beta_population}
    \beta_{cold(hot)} = \frac{1}{h\sqrt{\nu^{2}_{cold(hot)}+ (\frac{\Tilde{\omega}}{2\pi}})^2} \ln \left ( \frac{1- p^{+}_{cold(hot)}}{p^{+}_{cold(hot)}}     \right )
\end{equation}
where $p^{+}_{cold(hot)}=\bra{\Psi^{cold(hot)}_{+}}\rho^{th}_{in}(\rho^{th})\ket{\Psi^{cold(hot)}_{+}}$. When the population of high energy states exceeds the population of low-energy states, the effective temperature becomes negative. From Eq. (\ref{Equ:relation_beta_population}), we get that $p^{+}_{cold(hot)} \in [0, 0.5)$ leads to $\beta_{cold(hot)}$ to be positive while $p^{+}_{cold(hot)} \in [0.5, 1.0)$, thereby giving $\beta_{cold(hot)}$ negative. It was already shown that QOE can be efficient, when $\beta_{cold(hot)}$ takes negative values \cite{Assis_PRL_2019}. In our analysis, we vary $p^{+}_{hot}$ within \(0.5\) and \(1.0\) by fixing $p^{+}_{cold} =0.261$.

Let us first compute the partition function,
\begin{eqnarray}
    Z_{cold(hot)}= && \Tr[e^{-\beta_{cold(hot)}H_{cold(hot)}}] \nonumber \\=
     && 2\cosh \left ( \beta_{cold(hot)} E_{cold(hot)}  \right ). \nonumber \\
     && 
\end{eqnarray}
Now to derive the QOE efficiency($\eta$), we have to calculate 
the average net work $\langle W \rangle $ performed by the QOE and then the average heat $\langle Q_{hot} \rangle $. Lets calculate them with the constraints $\beta_{cold} >0 $ and $\beta_{hot} < 0 ~ (\beta_{hot} = - |\beta_{hot}|)$. The detail calculations are done in Appendix (\ref{App}). 
The net work of the cycle occurs during \(1 \rightarrow 2\) and  \(3 \rightarrow 4\).  Hence it  is defined as  
\begin{eqnarray}
\label{Eq:work}
    \langle W \rangle = && \langle W_{1\rightarrow2} \rangle + \langle W_{3 \rightarrow 4} \rangle \nonumber \\
    = && \Tr[\rho_{exp} H_{hot}] - \Tr[\rho^{th}_{in} H_{cold}] + \Tr[\rho_{comp} H_{cold}] \nonumber \\
    && - \Tr[\rho^{th} H_{hot}] \nonumber \\
    = && \Xi_{1} + \xi ~ \Xi_{2} 
\end{eqnarray}
where 
$$\Xi_{1} = (E_{cold}- E_{hot}) \left[ \tanh(\beta_{cold}E_{cold}) + \tanh(|\beta_{hot}|E_{hot})   \right ]$$
and 
$$\Xi_{2} = 2\left[ E_{hot} \tanh(\beta_{cold}E_{cold}) - E_{cold} \tanh(|\beta_{hot}|E_{hot})   \right ]$$
while the heat exchanged between the working medium and the hot as well cold reservoirs happens during other steps since they cannot take place simultaneously, i.e., during \(3 \rightarrow 2\) and \(4 \rightarrow 1\). They are respectively given by
\begin{eqnarray}
    \langle Q_{hot} \rangle = && \langle Q_{3 \rightarrow 2} \rangle \nonumber \\
    = && \Tr[\rho^{th} H_{hot}] - \Tr[\rho_{exp} H_{hot}] \nonumber \\
    =&& \Pi_{1} -\xi ~ \Pi_{2}
\end{eqnarray}
where 
$$
\Pi_{1}= E_{hot}\left[ \tanh(\beta_{cold}E_{cold})+ \tanh(|\beta_{hot}|E_{hot})  \right ]
$$
and  
$$
\Pi_{2} = 2 \left[ E_{hot} \tanh(\beta_{cold}E_{cold} )  \right ]
$$
In addition,
\begin{eqnarray}
    \langle Q_{cold} \rangle = && \langle Q_{4 \rightarrow 1} \rangle \nonumber \\
    = && \Tr[\rho^{th}_{in} H_{cold}] - \Tr[\rho_{comp} H_{cold}] \nonumber \\
    = && \Pi_{3} +\xi~ \Pi_{4}
\end{eqnarray}
where 
$$
\Pi_{3}= -E_{cold}\left[ \tanh(\beta_{cold}E_{cold})+ \tanh(|\beta_{hot}|E_{hot})  \right ]
$$
and 
$$
\Pi_{4} = 2 \left[ E_{cold} \tanh(|\beta_{hot}|E_{hot} )  \right ].
$$
Before we calculate the efficiency, we can note that the net work depends on transition probability $\xi$ in Eq. (\ref{Eq:work}). 
It is evident that when one of the thermal reservoirs is at a negative temperature, the transition probability $\xi$ can contribute to an increase in the extracted net work $\langle W \rangle$. Now $\xi$ serves as an adiabaticity parameter, indicating the speed at which the expansion and compression stages are performed. The relationship between the transition probability $\xi$ and the driving time $\tau$ is illustrated in Fig. \ref{fig:transitin_prob}. The study shows that the transition probability $\xi$ gets more value with less driving time. That means if the other QOHE parameters are properly adjusted, the faster the expansion and compression processes are performed the greater the contribution of this parameter $\xi$ to the extracted net work $\langle W \rangle$. Consequently, this should also lead to an increase in the efficiency $\eta$. Now 
by precisely adjusting the strength of the magnetic field in the \(z\)-direction, we can effectively enhance the transition probability further.  
Therefore, we can expect a better efficiency if we choose $\xi$ with  proper values of $\mathbf{g}$  (see Fig. \ref{fig:transitin_prob})  which is more than the one with  $\mathbf{g}=0$, 
although the impact is not ubiquitous. Note, however, that the transition probability $\xi$ is not the only parameter influencing $\langle W \rangle$ and $\langle Q_{hot} \rangle$. There are other parameters like the direction of the additional magnetic field \(\mathbf{g}\) that can have a greater impact when manipulated or when their signs are changed. Therefore, we cannot straightforwardly attribute the increase in $\xi$ as the sole reason for enhancing $\langle W \rangle$ or reducing $\langle Q_{hot} \rangle$, which, in turn, leads to an ultimate increase in the system's efficiency, $\eta$. The efficiency in terms of net work and heat can be computed analytically (see Appendix \ref{App}) as  
\begin{eqnarray}
\eta = && - \frac{\langle W \rangle }{\langle Q_{hot} \rangle } \nonumber \\
= && 1 - \frac{E_{cold}}{E_{hot}}  \left( \frac{1-2\xi\mathcal{F}}{1-2\xi \mathcal{G}}   \right) \nonumber \\ 
\end{eqnarray}
Here
\begin{eqnarray}
 \hspace{0.5in} \mathcal{F} = 
 \frac{\tanh\left(|\beta_{hot}|E_{hot} \right)} {\tanh\left(\beta_{cold}E_{cold}\right)+ \tanh\left(|\beta_{hot}|E_{hot} \right)  }, 
\end{eqnarray}
and 
\begin{eqnarray}
 \mathcal{G} =   \frac{\tanh\left(\beta_{cold}E_{cold} \right)} {\tanh\left(\beta_{cold}E_{cold}\right)+ \tanh\left(|\beta_{hot}|E_{hot} \right)  }. 
\end{eqnarray}
As seen in Fig. \ref{fig:transitin_prob}, we realize that when the driving time is moderate, $\xi$ gives higher value with  $\mathbf{g} > 0$ than that of the system having $\mathbf{g}=0$.

\begin{figure*}[htb!]
\includegraphics[width=\linewidth ]{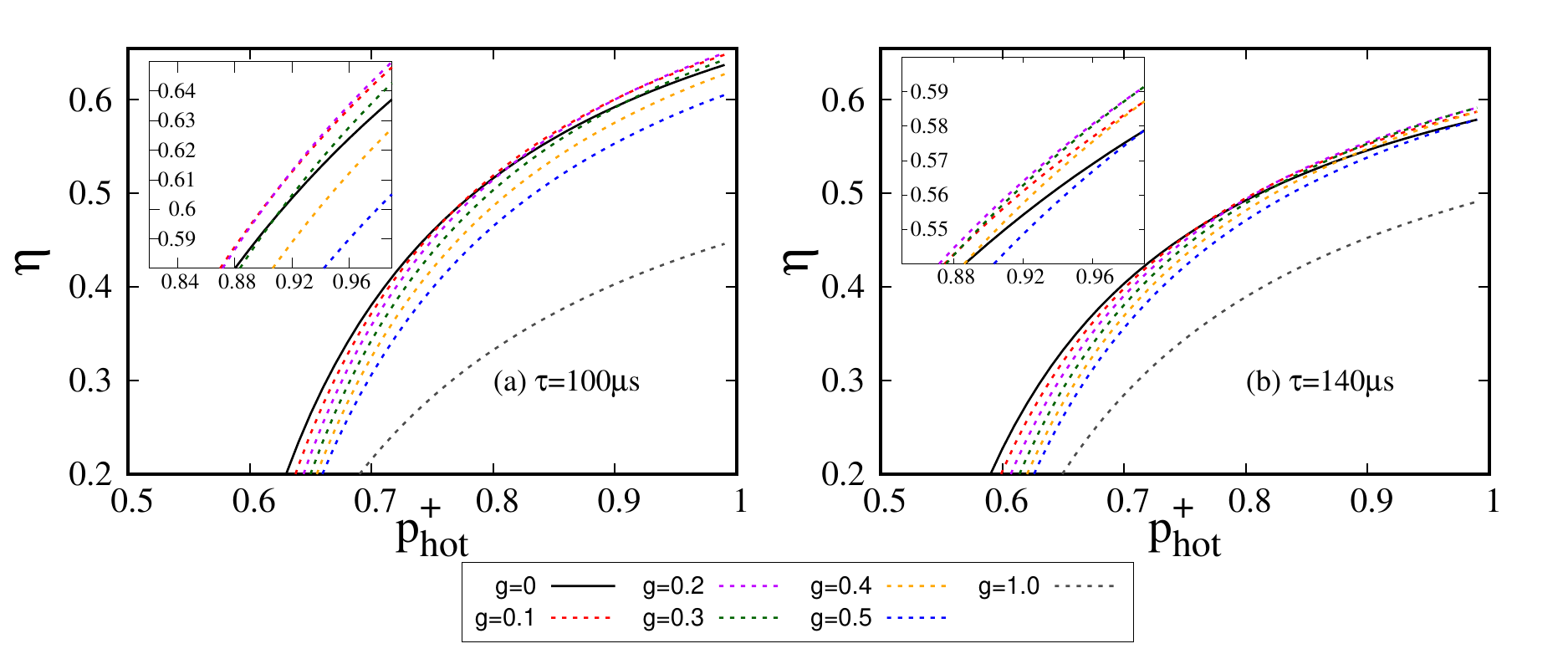}
\includegraphics[width=\linewidth ]{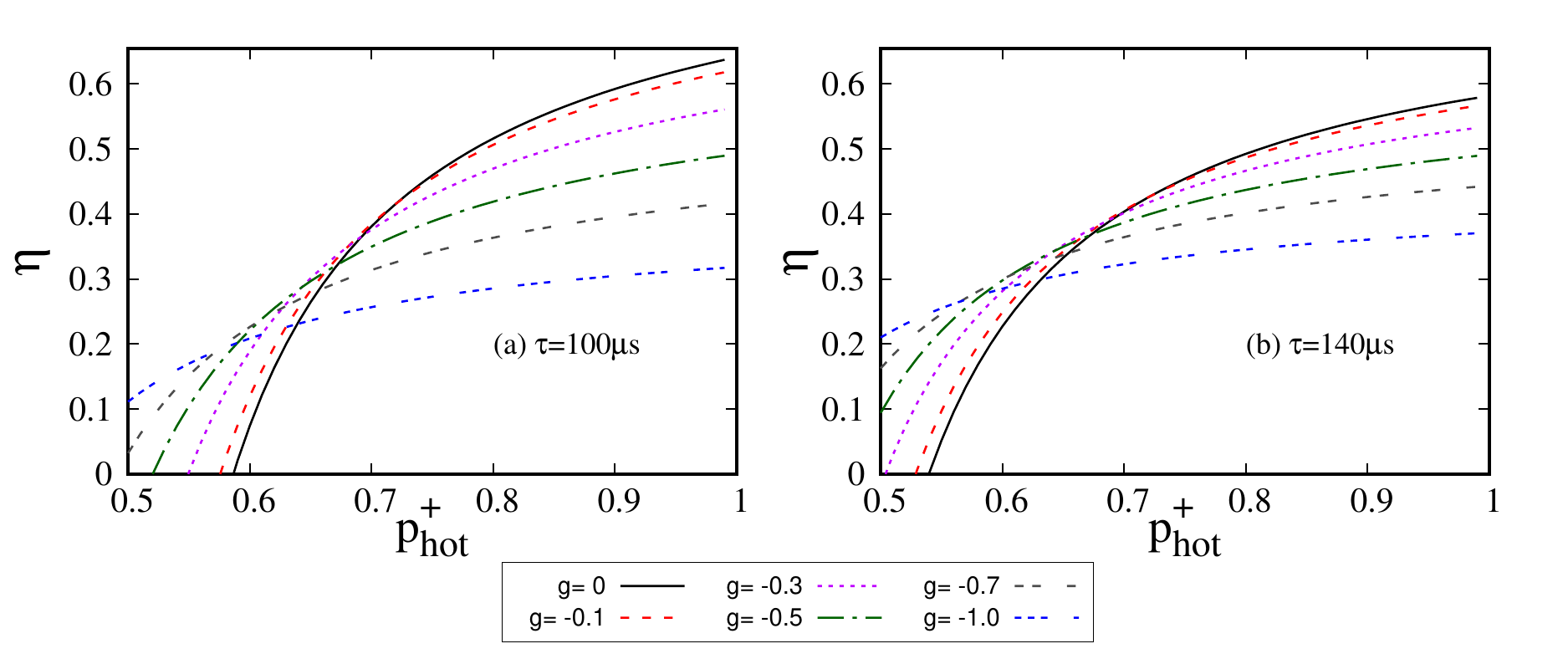}
\caption{ Efficiency, $\eta$, (ordinate) versus the population  of the excited state, $p^{+}_{hot}$ (abscissa) for a fixed driving time $\tau$.
Solid lines represent \(\mathbf{g}=0\) and all dotted lines are for \(\mathbf{g}>0\) and \(\mathbf{g}< 0\). 
We take $p^{+}_{cold}=0.261$ and vary $p^{+}_{hot}$ within $[0.5,1.0)$. All the other parameters are same as in Fig. \ref{fig:transitin_prob}.
(left) Driving time is fixed to \(\tau =100 \mu s\) and (right) it is \(\tau =140 \mu s\). 
The maximum efficiency is achieved at $\mathbf{g}=0.2$ for $\tau=100 \mu s$ while it  is $\mathbf{g}=0.3$ when $\tau=140 \mu s$. All the axes are dimensionless. }
\label{fig:eff_ph}
\end{figure*}

Let us concentrate on $\tau \in [100, 200] \, \text{in} \, \mu s$. Note that the range of \(\tau\) is also chosen for demonstration in the NMR experiment of QOE   \cite{Assis_PRL_2019} which implies that such driving time is accessible in the laboratory. In this regime, we observe that when $p^{+}_{hot}$ is in the neighbourhood of unity, the efficiency of QOE with system having $\mathbf{g}>0$ is higher than that of the system without the magnetic field. The small driving time leads to a higher range of $p^{+}_{hot}$ values for which 
$
\eta(\mathbf{g}> 0) > \eta(\mathbf{g}=0) $ (as depicted in Fig. \ref{fig:eff_ph}).

Interestingly, introduction of magnetic field  is not  ubiquitous to obtain a high efficiency in QOE. Specifically, there exists a critical strength of the magnetic field, $\mathbf{g}_{c}$, above which the efficiency goes below the one obtained with $\mathbf{g}=0$, when all the parameters are fixed to a certain value. For a fixed $\tau$, and $p^{+}_{cold}$, there exists a lower bound of $p^{+}_{hot}$ above which high $\eta(\mathbf{g}>0)$ is obtained. More precisely, we notice that 
$$ \eta(\tau=100 \mu s,\mathbf{g}=0.2) - \eta(\tau=100 \mu s,\mathbf{g}=0) =0.012, $$ 
where we fix $p^{+}_{cold}= 0.261$ and $p^{+}_{hot}= 0.99$ while 
$$ \eta(\tau=200 \mu s,\mathbf{g}=0.3) - \eta(\tau=100 \mu s,\mathbf{g}=0) =0.007 $$
for the same set of values of  $p^{+}_{cold}$ and $p^{+}_{hot}$.
For $\tau=140 \mu s$, the efficiency advantage we get is $0.012$ for $g=0.3$ and the other parameters $p^{+}_{cold(hot)}$ are same as above. 
When
 $\xi =0$, it represents the conventional heat engine with efficiency $\eta = 1 - (E_{cold}/E_{hot})$. When the expansion and compression strokes follow ideal adiabatic process, i.e., $\xi=0$, $\eta$ tends to $\eta_{Otto}$. The condition for which we can get \( \eta\geq\eta_{Otto}\) is given by (see Appendix \ref{App})
\begin{equation}
    |\beta_{hot}| \sqrt{4\pi^{2}\nu^{2}_{hot}+\Tilde{\omega}^{2}}   \geq  \beta_{cold} \sqrt{4\pi^{2}\nu^{2}_{cold}+\Tilde{\omega}^{2}}
\end{equation}

If we change the direction of the additional magnetic field in the \(z\)-direction i.e. $\mathbf{g} <0$,  we can highlight another important aspect  of our model. In particular, the engine  with $\mathbf{g} <0$ operates within an extended operational domain, reaching into temperature ranges where the  effective negative temperature-based quantum Otto engine operating only on the rotational magnetic field in \((x,y)\)-plane cannot  achieve. Furthermore, this extended operational domain of the engine increases with the increase of  $\textbf{g}$  (see Fig. \ref{fig:eff_ph}). Let us take $\mathbf{g} <0$ and ask whether the benefit in efficiency  reported above with \(\mathbf{g}>0\) persists.
The answer is affirmative. We observe that with $\mathbf{g}<0$,  $ \eta(\mathbf{g}<0) > \eta(\mathbf{g}=0) $ upto certain values of $\mathbf{g}$. However, unlike $\mathbf{g}>0$, where the advantage is obtained when $p^{+}_{hot} \approx 1$, we find that the enhancement of $\eta$ is found, when  $p^{+}_{hot}$ is close to $0.5$ with the same values of \(p^{+}_{cold}\). More precisely, if we take 
$$ \eta(\tau=100 \mu s,\mathbf{g}=-0.5) - \eta(\tau=100 \mu s,\mathbf{g}=0) =0.146, $$ 
where we fix $p^{+}_{cold}= 0.261$ and $p^{+}_{hot}= 0.6$ while 
$$ \eta(\tau=140 \mu s,\mathbf{g}=-0.5) - \eta(\tau=140 \mu s,\mathbf{g}=0) =0.07 $$
for the same set of values of  $p^{+}_{cold}$ and $p^{+}_{hot}$. There exists a critical effective negative temperature or in terms of the population of the excited state $p^{+}_{hot}$ below which we can get the advantage of the  efficiency and the critical temperature depends on the negative values of $\textbf{g}$. As the negative $\textbf{g}$ value rises, the critical negative temperature lessens according to that. Interestingly, by the help of the additional magnetic field in \(z\)-direction, the protocol can operate as an  engine within the full range of effective negative temperature i.e. $p^{+}_{hot} \in [0.5, 1.0)$ which is not possible without the additional magnetic field. 

We will now try to link the the advantage obtained with the addition of magnetic field in each stroke with coherence of the system.

\subsection{Coherence: An indicator of enhanced performance in QOE}

Let us investigate the behavior of coherence for the expansion and compression stroke in the Otto cycle. It was indicated that the performance of four-stroke Otto cycle can be linked to coherence \cite{Feldmann_PRE_2006, Rezek_NJP_2006, Feldmann_PRE_2012, Thomas_EPJB_2014, Luis_PRE_2015, Santos_npj_2019, Francica_PRE_2019}. The Hamiltonian $H_{exp}(t)$ and $H_{comp}(t)$ do not commute with itself at different times, so these non-commuting Hamiltonian generates coherence in the expansion and compression strokes of the QOE while it vanishes during thermalization. Here we assume that coherence is computed in the energy eigenbasis of $H_{cold}$ and $H_{hot}$. 

We quantify the coherence by $l_{1}$-norm measure which is defined as the sum of the absolute values of the off diagonal elements of a given state which is written in the eigenbasis of $H_{cold}$ and $H_{hot}$, i.e., $C_{l_1} (\rho) = \sum_{i \neq j} |\rho_{ij}| $ \cite{Streltsov_RMP_2017}. 
One of our goal is to examine how the coherence is affected by introducing $\Tilde{\omega}$ in the driving Hamiltonian.  To study this, we calculate coherence with respect to \(\tau\) for different $\mathbf{g}$ values.
\begin{figure*}
\includegraphics[width=\linewidth ]{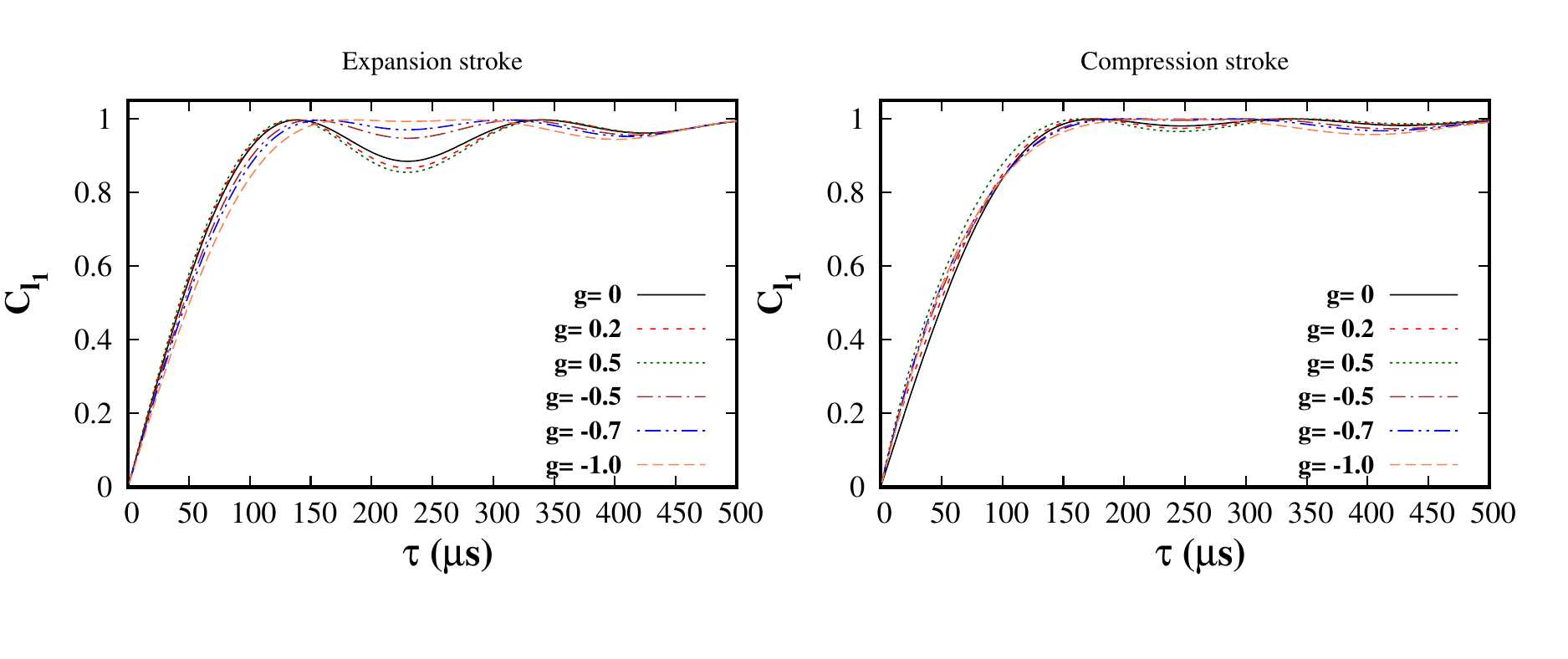}
\caption{$l_1$-norm of coherence ($C_{l_{1}}$) (vertical axis) with  driving time $\tau$ (horizontal axis) for different $\mathbf{g}$ values. The unit of the horizontal axis is  $\mu s$  and the vertical axis is dimensionless. We calculate the coherence for $p^{+}_{hot}=0.9$. All the other specifications are same as in Fig. \ref{fig:eff_ph}.   }
\label{fig:exp-comp_stroke_coh}
\end{figure*}
Specifically, during the expansion and the compression strokes, we study the trends of coherence measure with the variation of the driving time ($\tau$) for different strengths and direction in the \(z\)-directional magnetic field, $\mathbf{g}$.  We can observe that there exits a range of $\mathbf{g}$ for which the coherence is more than that obtained $\mathbf{g}=0$ (see Fig. \ref{fig:exp-comp_stroke_coh}). Since the states are different for the expansion and compression strokes, the coherence is not equal. We observe that, up to a certain driving time, the coherence generated by manipulating $\mathbf{g}$ is higher than that produced without the magnetic field in the \(z\)-direction. The graph's nature changes when the direction of 
of $\mathbf{g}$ is altered.
However, this can only be observed in the measure of the expansion stroke.
For positive $\mathbf{g}$ values, the system exhibits greater coherence initially (for small $\tau$ values), although it is smaller compared to the engine with $\mathbf{g}=0$  as time progresses. Conversely, for negative $\mathbf{g}$ values, the pattern is reversed. This difference is also reflected in the efficiency study. However,  we cannot find a close relation between $C_{l_{1}}$ and the efficiency of QOE, $\eta$.  

 \begin{figure*}[htb!]
\includegraphics[width=\linewidth ]{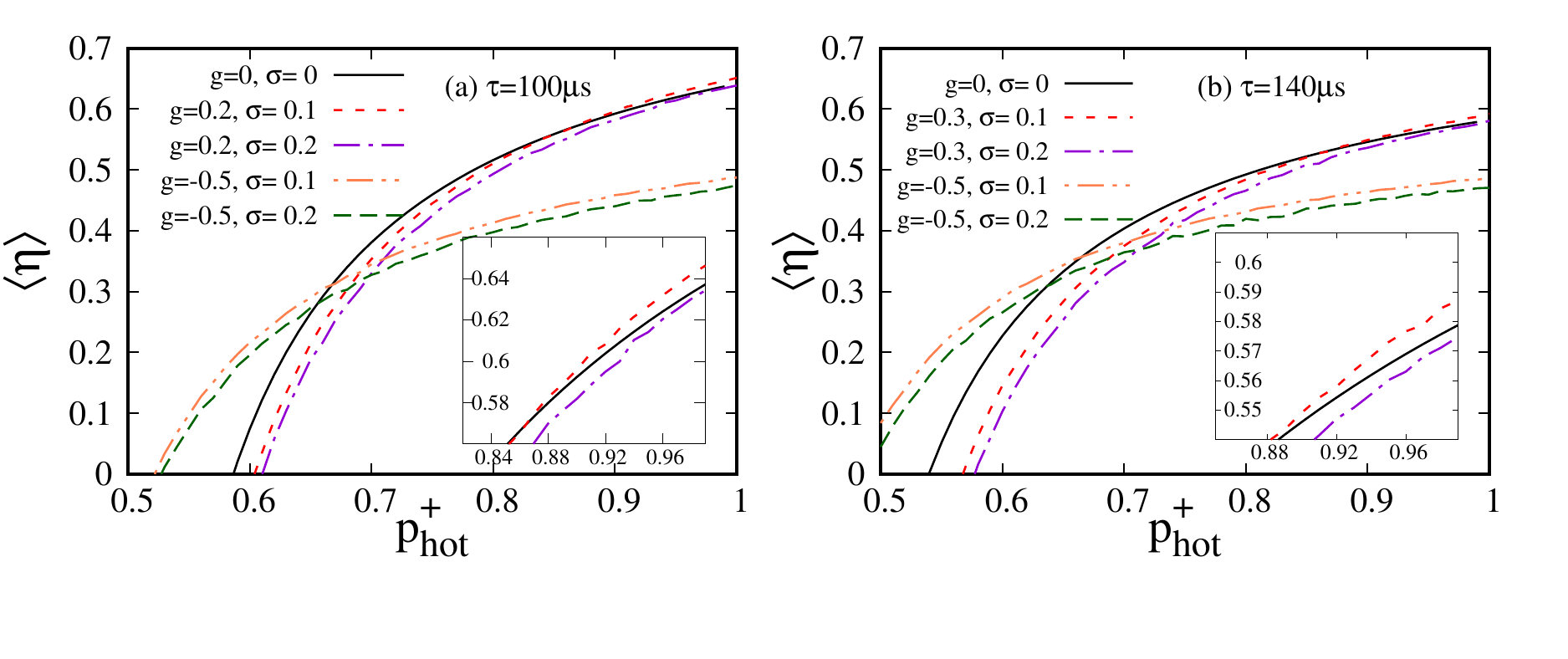}
\caption{Quenched average efficiency, denoted as $\langle\eta\rangle$ (ordinate) against  the population  of the excited state, $p^{+}_{hot}$ (abscissa) for two different values of $\mathbf{g}$.  \(\delta_i\) (\(i=1,2\)) involved in \((1 + \delta_1)\nu_{cold}\) and \((1+\delta_2)\nu_{hot}\) of the QOE  are chosen randomly from  two Gaussian distributions with zero mean and the same standard deviation, $\sigma$ for a fixed driving time $\tau$. 
(left) We compare  \(\mathbf{g}=0.2\) and \(\mathbf{g}=-0.5\) by varying the strength of the disorder, \(\sigma\) with the QOE having \(\mathbf{g} =0\) without disorder. (right) The same comparison is performed with \(\mathbf{g} =0.3\) and \(\mathbf{g} =-0.5\). Choices of \(\mathbf{g}\) is due to the observations in Fig. \ref{fig:eff_ph}. 
We take all the others parameters are same as in Fig. \ref{fig:eff_ph}.
All the axes are dimensionless.    }
\label{fig:Gauss_dist_compare}
\end{figure*}

\section{Robustness of Otto engine against impurities}
\label{disorder}

During implementation of any quantum devices, some imperfection or impurities or noise is inevitable, thereby creating hindrances in the performance \cite{Maciej_AIP_2007, Ahufinger_PRA_2005, Tanoy_PRA_2022, Ahana_PRA_2020, Srijon_PRA_2020, Maciej_Nature_2010, breuer2002, rivas2012}. To study such scenario, decoherence or disorder can be introduced during several steps of the engines. For example, noise can effect during the preparation of the initial state \cite{Andolina19, Srijonfast21}, or  imperfect unitary operator \cite{Pritam_PRA_2022} can also reduce the performance of the engine.  Generally, disorders are often present in the engine's control and in the measurement process \cite{Kosloff_2014}. Due to impurities in the Hamiltonian, the quantum coherence may be disturbed, thereby leading to a change in efficiency. To attenuate the effects of disorder on the quantum engine, there are some approaches that can be taken like quantum error correction \cite{E.Knill_Nature_2005, Terhal_RMP_2015, Iulia_Nature_2020}, dynamical decoupling 
 \cite{Viola_PRL_1999, Alvarez_PRL_2011}, and robust control methods \cite{Dong_Robust_book, shermer_2023}. Further, the current experimental developments also give rise to the possibility to probe properties of such disordered models where disorder can be added in the system.

Let us now check the effect of
disorder on the design of the four-stroke engine as the disorder plays a vital role in the performance of the quantum device. 
Hence it is natural to ask how these types of engines react to impurities or disorders which has never been addressed before. The impact of the disorder can be two-fold: it may reduce the efficiency by creating extra entropy in the system while in disordered engine,  quantum coherence may raise in some parameter regimes,  thereby leading to a favorable effect on the system output \cite{Tanoy_PRA_2022, Ujjal_JOP_2022, Ahana_PRA_2020, Srijon_PRA_2020, Anindita_EPL_2019, Debasis_PRE_2016, Aharony_PRB_1978, Villain_JPF_1980, Hide_PRL_2009, Fujii_PRA_2010, Prabhu_PRA_2011, Villa_PRE_2014, Niederberger_PRL_2008, Niederberger_EPL_2009, Patrick_RMP_1985, Maciej_Nature_2010, Debasis_PRE_2016, Anindita_PRB_2016, Anindita_PRB_2017, Tong_PRA_2010}. 
To introduce disorder in the model, a single or several parameters involved in the model are  chosen randomly, and one then preforms averaging of the quantity of interest which, in this case, is efficiency in over large number of configurations - quenched averaging. In this  scenario, it is assumed that the observation time of some parameters is much smaller than the time taken by the same set of parameters to reach the equilibrium. Quenched disorder typically creates a energy landscape with many local minima and it confines the system in a metastable state, thereby remaining in one of the local minima for a long time \cite{Sethna_book, Bouchaud_JPI_1992, Fisher_PRB_1988, Binder_RMP_1986, Suzuki_book, subir_sachdev_book, BKChakrabarti_book, MMezard_book, DChowdhury_book}.

 We here incorporate disorder in the frequency of the ideal system, $\nu_{cold} $ and $\nu_{hot}$ as $(1+\delta_1)\nu_{cold}$ and $(1+\delta_2)\nu_{hot}$. Here $\delta_{1(2)}$ 
 is chosen from  two different Gaussian distributions having same vanishing mean and  standard deviation $\sigma$ which can be called strength of the disorder. The probability density function for Gaussian  disorder with vanishing mean and standard deviation, \(\sigma\) is given by
 \(   P(\delta_i) = \frac{1}{\sigma \sqrt{2\pi}} e^{-\frac{1}{2}(\frac{\delta_i}{\sigma})^2}, ~ ~ -\infty  \leq \delta \leq \infty\).
 The quenched averaged efficiency for a given strength of disorder, \(\sigma\) can be written as
$$
\langle \eta(\sigma) \rangle = \int P(\delta_1) P(\delta_2) \eta \left ( \nu_{cold}(1+\delta_1), \nu_{hot}(1+\delta_2) \right )d\delta_1 \delta_2.
$$

We have already shown that in presence of magnetic field in the z-direction, the efficiency of the four-stroke machine can be improved. We now want to find out whether even in presence of disorder, the advantage persists or not. For comparison, we take the system with $\mathbf{g}=0$ without disorder. For a fixed driving time, a fixed $\mathbf{g}>0$ or a fixed $\mathbf{g}< 0$ and $p^{+}_{cold}$, we compute the quenched average efficiency, $\langle \eta \rangle $ by varying $p^{+}_{hot}$ for different values of disorder strength. 
We have taken $10^{4}$ realizations to compute the quenched averaged efficiency. We observe that the efficiency slightly decreases in presence of impurities in $\nu_{hot}$ and $\nu_{cold}$. However,  we find that if the strength of the disorder is moderate, $\langle \eta \rangle $ is higher than that can be obtained by the system with $\mathbf{g}=0$ (as shown in Fig. \ref{fig:Gauss_dist_compare}).

Again, there is a trade-off relation between $\sigma$ and the driving time $\tau$. Surprisingly, the robustness of this QOE based on effective negative temperature increases with the increase of \(p^+_{hot}\) (in Fig. \ref{fig:Gauss_dist_compare}, compare \(\langle \eta\rangle\) when \(p^+_{hot} \sim 0.5\) and \(p^+_{hot}\sim 1.0\) ). 
We emphasis here that the four-stroke quantum Otto engine can be shown to be robust against disorder, when parameters are chosen both from Gaussian and uniform distributions, given by  (\(
    P(\delta) = \frac{1}{\sigma} , ~ ~ \text{when}  -\frac{\sigma}{2}  \leq \delta \leq \frac{\sigma}{2}\)
and \(   = 0,  \,\text{otherwise}\) 
(for illustration, we only consider Gaussian disorder in Fig. \ref{fig:Gauss_dist_compare}).

\section{Conclusion}
\label{conc}

The performance of thermal machines such as heat engines, refrigerators, and batteries can be significantly improved by incorporating quantum principles into their design. 
This research article centers around the study of a four-stroke quantum Otto engine, which operates between two thermal reservoirs -- one is characterized by a positive spin temperature while the other one features an effective negative temperature, indicating a population inversion in the spin system. Remarkably, it was demonstrated in laboratories that this type of engine exhibits enhanced efficiency compared to engines operating solely with reservoirs at positive temperatures \cite{Assis_PRL_2019}.

In our study, we employed a four-stroke quantum Otto cycle protocol, in which the driving Hamiltonian is considered in its general form. Specifically, a rotating magnetic field in the \((x, y)\)-plane, and an independent magnetic field in the \(z\)-direction with varying strengths and the direction are involved in the design. By adopting this general form of the driving Hamiltonian, the capability to manipulate the strength of the magnetic fields individually in both directions can be achieved, thereby providing a possibility to create different classes of dynamical states, suitable for QOE. Through our investigation, we found that this is indeed the case. In particular, we uncovered that by precisely adjusting the strength and the direction of the magnetic field in the \(z\)-direction and manipulating other relevant system parameters, we can significantly enhance the efficiency of the engine in comparison to the scenario where there is no magnetic field present in the \(z\)-direction. 
Additionally, a pivotal facet of our model rests in its capability to operate within an extended operational domain. It can achieve temperature ranges where the  effective negative temperature-based quantum Otto engine functioning solely on the rotational magnetic fields within the \((x,y)\) plane, recently constructed in laboratories is unable to operate. Fascinatingly, with the aid of an additional magnetic field oriented in the \(z\)-direction, the protocol is able to function as an engine across the entire spectrum of the effective negative temperatures.
The advantageous aspect is that our protocol is also feasible for experimental implementation which we also ensure by choosing certain system parameters from recent experimental demonstration \cite{Assis_PRL_2019}.

For the analytical treatment, the unitary evolution of the system is formulated in a rotating basis. Using this approach, we calculated the transition probability as a function of driving time with the variation in the strength of the magnetic field in the \(z\)-direction. Notably, we observed that the transition probability exhibits an increment for specific values of the strength of the magnetic field, leading to the enhanced efficiency in the quantum engine. To further analyze the engine's performance, we analytically expressed the total work and average heat, allowing us to derive a closed-form expression for the efficiency of the engine. Interestingly, we found that the increase in the strength of the additional magnetic field  does not universally enhance the efficiency. Instead, there exists a critical strength for the magnetic field at a fixed driving time which leads to the improvement in the engine. Additionally, we observed that at very high driving times, such an  advantage in efficiency diminishes. We argued that the generation of coherence during the expansion and compression strokes, along with the transition probability for a moderate driving time, is linked to the increased efficiency of the proposed quantum Otto engine.
The results obtained here indicate that the performance of the quantum heat engines can be improved even by locally tunable properties like the coherence of the system.

In the realization of quantum devices, imperfections or impurities inevitably arise during their construction or operation. Given the inherent presence of disorder, it becomes crucial to estimate the robustness of these devices against such imperfections in order to achieve precision and accuracy.
In our study, we specifically investigated the impact of the disorder on the frequencies associated with all the strokes of the quantum engine. We illustrated that even when these parameters are randomly chosen from Gaussian or uniform distributions, the quenched averaged efficiency of the system can still surpass that of the engine without a constant magnetic field, thereby demonstrating the robustness of the presented quantum Otto engine against impurities.

This work presents a promising avenue for future research. Firstly, the observed increase in efficiency opens up possibilities for exploring similar advancements in other thermodynamic engines by manipulating locally addressable system parameters. Additionally, the demonstrated robustness against impurities holds a great promise for building quantum engines including the implementation of effective negative temperature in laboratories with other physical substrates.

\section{Acknowledgment}
We acknowledge the support from the Interdisciplinary Cyber-Physical Systems (ICPS) program of the Department of Science and Technology (DST), India, Grant No.: DST/ICPS/QuST/Theme- 1/2019/23. 

\newpage
\begin{widetext}
\appendix
\section{Computation of efficiency via net work and heat }
\label{App}
The net work of the cycle in the four-stroke QOE can be defined as
\begin{eqnarray}
    \langle W \rangle = && \langle W_{1\rightarrow2} \rangle + \langle W_{3 \rightarrow 4} \rangle \nonumber \\
    = && \Tr[\rho_{exp} H_{hot}] - \Tr[\rho^{th}_{in} H_{cold}] + \Tr[\rho_{comp} H_{cold}] \nonumber \\
    && - \Tr[\rho^{th} H_{hot}].\nonumber
\end{eqnarray}
On the other hand,  the heats exchanged between the working medium and the hot reservoir can be written as
\begin{eqnarray}
    \langle Q_{hot} \rangle = && \langle Q_{3 \rightarrow 2} \rangle \nonumber \\
    = && \Tr[\rho^{th} H_{hot}] - \Tr[\rho_{exp} H_{hot}]. 
\end{eqnarray}
When the heat exchange occurs between the medium and cold reservoir, the above quantity takes the form as
\begin{eqnarray}
    \langle Q_{cold} \rangle = && \langle Q_{4 \rightarrow 1} \rangle \nonumber \\
    = && \Tr[\rho^{th}_{in} H_{cold}] - \Tr[\rho_{comp} H_{cold}].  
\end{eqnarray}
Let us analytically compute the each quantity involved in the definition of heat and net work. 
\begin{eqnarray}
     \Tr[\rho_{exp} H_{hot}] = 
     && \Big[ - \frac{h}{4\pi} \sqrt{4\pi^{2}\nu^{2}_{hot}+\Tilde{\omega}^{2} }~~
   \tanh\left(  \beta_{cold}  \frac{h}{4\pi} \sqrt{4\pi^{2}\nu^{2}_{cold}+\Tilde{\omega}^{2} }
 \right )  \left( 1 - 2\xi \right ) \Big ], \nonumber \\
 = && - E_{hot} \tanh(\beta_{cold}E_{cold}) \left( 1-2\xi\right ),
\end{eqnarray}

\begin{eqnarray}
     \Tr[\rho_{comp} H_{cold}] = 
    && \Big[ - \frac{h}{4\pi} \sqrt{4\pi^{2}\nu^{2}_{cold}+\Tilde{\omega}^{2} }~~
   \tanh\left(  \beta_{hot}  \frac{h}{4\pi} \sqrt{4\pi^{2}\nu^{2}_{hot}+\Tilde{\omega}^{2} }
 \right )  \left( 1 - 2\xi \right ) \Big ], \nonumber \\
 = && - E_{cold} \tanh(\beta_{hot}E_{hot}) \left( 1-2\xi\right ),
\end{eqnarray}
and
\begin{eqnarray}
     \Tr[\rho^{th}_{in} H_{cold}] = 
    && \Big[ - \frac{h}{4\pi} \sqrt{4\pi^{2}\nu^{2}_{cold}+\Tilde{\omega}^{2} }~~
   \tanh\left(  \beta_{cold}  \frac{h}{4\pi} \sqrt{4\pi^{2}\nu^{2}_{cold}+\Tilde{\omega}^{2} }
 \right ) \Big ], \nonumber \\
 = && - E_{cold} \tanh(\beta_{cold}E_{cold}) ,
\end{eqnarray}
and
\begin{eqnarray}
 \Tr[\rho^{th} H_{hot}] = 
    && \Big[ - \frac{h}{4\pi} \sqrt{4\pi^{2}\nu^{2}_{hot}+\Tilde{\omega}^{2} }~~
   \tanh\left(  \beta_{hot}  \frac{h}{4\pi} \sqrt{4\pi^{2}\nu^{2}_{hot}+\Tilde{\omega}^{2} }
 \right )  \Big ], \nonumber \\
 = && - E_{hot} \tanh(\beta_{hot}E_{hot}) .
\end{eqnarray}
With the help of the above quantities, we can arrive at  the expression of the  $\langle W \rangle$,  $\langle Q_{hot} \rangle$ and  $\langle Q_{cold} \rangle$ as

\begin{eqnarray}
   \langle W \rangle = \left( E_{cold}-E_{hot} \right) && \Big[ \tanh(\beta_{cold}E_{cold})-\tanh(\beta_{hot}E_{hot}) \Big] + \nonumber \\ 
   && 2\xi \Big[ E_{hot}\tanh(\beta_{cold}E_{cold})+E_{cold} \tanh(\beta_{hot}E_{hot}) \Big ],
\end{eqnarray}

\begin{eqnarray}
     \langle Q_{hot} \rangle = E_{hot} \Big[ \tanh(\beta_{cold}E_{cold}) - \tanh(\beta_{hot}E_{hot}) \Big ] -2\xi \Big[  E_{hot} \tanh(\beta_{cold}E_{cold})  \Big ],
\end{eqnarray}
and 
\begin{eqnarray}
     \langle Q_{cold} \rangle = -E_{cold} \Big[ \tanh(\beta_{cold}E_{cold}) - \tanh(\beta_{hot}E_{hot}) \Big ] -2\xi \Big[  E_{cold} \tanh(\beta_{hot}E_{hot})  \Big ].
\end{eqnarray}
Now by considering $\beta_{cold}>0$ and $\beta_{hot}<0 ~(\beta_{hot}=-|\beta_{hot}|)$, the above expressions reduces to

\begin{eqnarray}
   \langle W \rangle = \left( E_{cold}-E_{hot} \right) && \Big[ \tanh(\beta_{cold}E_{cold})+\tanh(|\beta_{hot}|E_{hot}) \Big] + \nonumber \\ 
   && 2\xi \Big[ E_{hot}\tanh(\beta_{cold}E_{cold})-E_{cold} \tanh(|\beta_{hot}|E_{hot}) \Big ],
\end{eqnarray}

\begin{eqnarray}
     \langle Q_{hot} \rangle = E_{hot} \Big[ \tanh(\beta_{cold}E_{cold}) + \tanh(|\beta_{hot}|E_{hot}) \Big ] -2\xi \Big[  E_{hot} \tanh(\beta_{cold}E_{cold})  \Big ],
\end{eqnarray}
and 
\begin{eqnarray}
     \langle Q_{cold} \rangle = -E_{cold} \Big[ \tanh(\beta_{cold}E_{cold}) + \tanh(|\beta_{hot}|E_{hot}) \Big ] + 2\xi \Big[  E_{cold} \tanh(|\beta_{hot}|E_{hot})  \Big ].
\end{eqnarray}

With the help of the above quantities, we can arrive at  the expression of the efficiency as
\begin{eqnarray}
\eta = && - \frac{\langle W \rangle }{\langle Q_{hot} \rangle } \nonumber \\
= && \frac{ \left( E_{cold}-E_{hot} \right) \Big[ \tanh(\beta_{cold}E_{cold})+\tanh(|\beta_{hot}|E_{hot}) \Big] + 2\xi \Big[ E_{hot}\tanh(\beta_{cold}E_{cold})-E_{cold} \tanh(|\beta_{hot}|E_{hot}) \Big ] }{ E_{hot} \Big[ \tanh(\beta_{cold}E_{cold}) + \tanh(|\beta_{hot}|E_{hot}) \Big ] -2\xi \Big[  E_{hot} \tanh(\beta_{cold}E_{cold})  \Big ]} \nonumber \\
= && 1 - \frac{E_{cold}}{E_{hot}} \left( \frac{1-2\xi\mathcal{F}}{1-2\xi \mathcal{G}}   \right).
\end{eqnarray}
 Here, let us define $F$ and $G$ as
 \begin{eqnarray}
 \mathcal{F} = &&
 \frac{\tanh\left(|\beta_{hot}|E_{hot} \right)} {\tanh\left(\beta_{cold}E_{cold}\right)+ \tanh\left(|\beta_{hot}|E_{hot} \right)  } ,
\end{eqnarray}
and 
\begin{eqnarray}
  \mathcal{G} = &&
 \frac{\tanh\left(\beta_{cold}E_{cold} \right)} {\tanh\left(\beta_{cold}E_{cold}\right)+ \tanh\left(|\beta_{hot}|E_{hot} \right)  } .
\end{eqnarray}


Let us now find out the condition for which we can get \( \eta\geq\eta_{Otto}\) in the form of $F$ and $G$, i.e. ,

\begin{eqnarray}
    1 - \frac{E_{cold}}{E_{hot}} \left( \frac{1-2\xi\mathcal{F}}{1-2\xi \mathcal{G}}   \right) & \geq & 1 - \frac{E_{cold}}{E_{hot}}   \nonumber \\
    \implies \left( \frac{1-2\xi\mathcal{F}}{1-2\xi \mathcal{G}}   \right)  & \leq & 1   \nonumber \\
  \implies F  & \geq & G  \nonumber \\
\implies \tanh\left( |\beta_{hot}| \frac{h}{4\pi} \sqrt{4\pi^{2}\nu^{2}_{hot}+\Tilde{\omega}^{2}} ~ \right) & \geq &  \tanh\left( \beta_{cold} \frac{h}{4\pi} \sqrt{4\pi^{2}\nu^{2}_{cold}+\Tilde{\omega}^{2}} ~ \right) \nonumber \\
  \implies  |\beta_{hot}| \sqrt{4\pi^{2}\nu^{2}_{hot}+\Tilde{\omega}^{2}}  & \geq & \beta_{cold} \sqrt{4\pi^{2}\nu^{2}_{cold}+\Tilde{\omega}^{2}} . 
\end{eqnarray}

\end{widetext}
\bibliographystyle{apsrev4-1}
\bibliography{Heat_Engine}
\end{document}